\documentstyle[preprint,prl,aps,epsf]{revtex} 
\input epsf
\newcommand{\bb}{\begin{equation}}
\newcommand{\en}{\end{equation}}    

\draft 
\begin{document}

\title{Binding of Oppositely Charged Membranes and Membrane Reorganization}
\author{A. W. C. Lau$\dag$ and P. Pincus$\dag\,\ddag$\\
$\dag$ Department of Physics,$\ddag$ Department of Materials,
and $\ddag$ Materials Research Laboratory,\\ 
University of California Santa Barbara, CA 93106-9530}
\date{\today}
\maketitle
\begin{abstract}
We consider the electrostatic interaction between two rigid 
membranes, with different surface charge densities of opposite sign, across 
an aqueous solution without added salt.  Exact solutions to the nonlinear 
Poisson-Boltzmann equation are obtained and their physical 
meaning discussed.  We also calculate the electrostatic contribution 
to the free energy and discuss the renormalization of the 
area per head group of the charged lipids arising from the 
Coulomb interaction.
\end{abstract}

\section{Introduction}
\label{sec:introduction}
Electrostatics often plays an important role in determining 
the structure of macromolecules in aqueous solution, 
{\em e.g.} polyelectrolytes, charged membranes, 
and charged colloidal particles.  For many physical systems, 
where charge densities on the surfaces are equal 
and of the same sign, the Poisson-Boltzmann equation provides quantitative
descriptions of their electrostatic interactions\cite{inter}.  
However, many biological processes involve 
charge densities on the surfaces that are not equal 
and sometimes even have opposite sign\cite{unequal}.  
Examples of this situation include protein association with DNA and 
membranes; the interaction between cationic
liposomes and negatively charged cell membranes\cite{biology}.
It also has significant biotechnological importance for the study of DNA 
association with artificial cationic liposomes\cite{gene}.  
In a recent experimental work on two oppositely charged membranes 
\cite{adhesion}, it is found that the membrane, due to its fluidity, 
adjusts its area per head in response to the electrostatic interaction.  
Similar conclusion has been reached by Radler {\it et. al.}\cite{DNA}
on DNA-cationic liposome complexes.  Motivated by these experiments,   
in the present paper we study a model of two oppositely 
charged membranes in electrostatic interaction and quantify how the area
per head renormalizes by using the Schulman-Montagne condition\cite{con}
for membrane self-assembly.

\begin{figure}[t]
\epsfxsize = 4in
\epsfysize = 4in
\centerline{\epsfbox{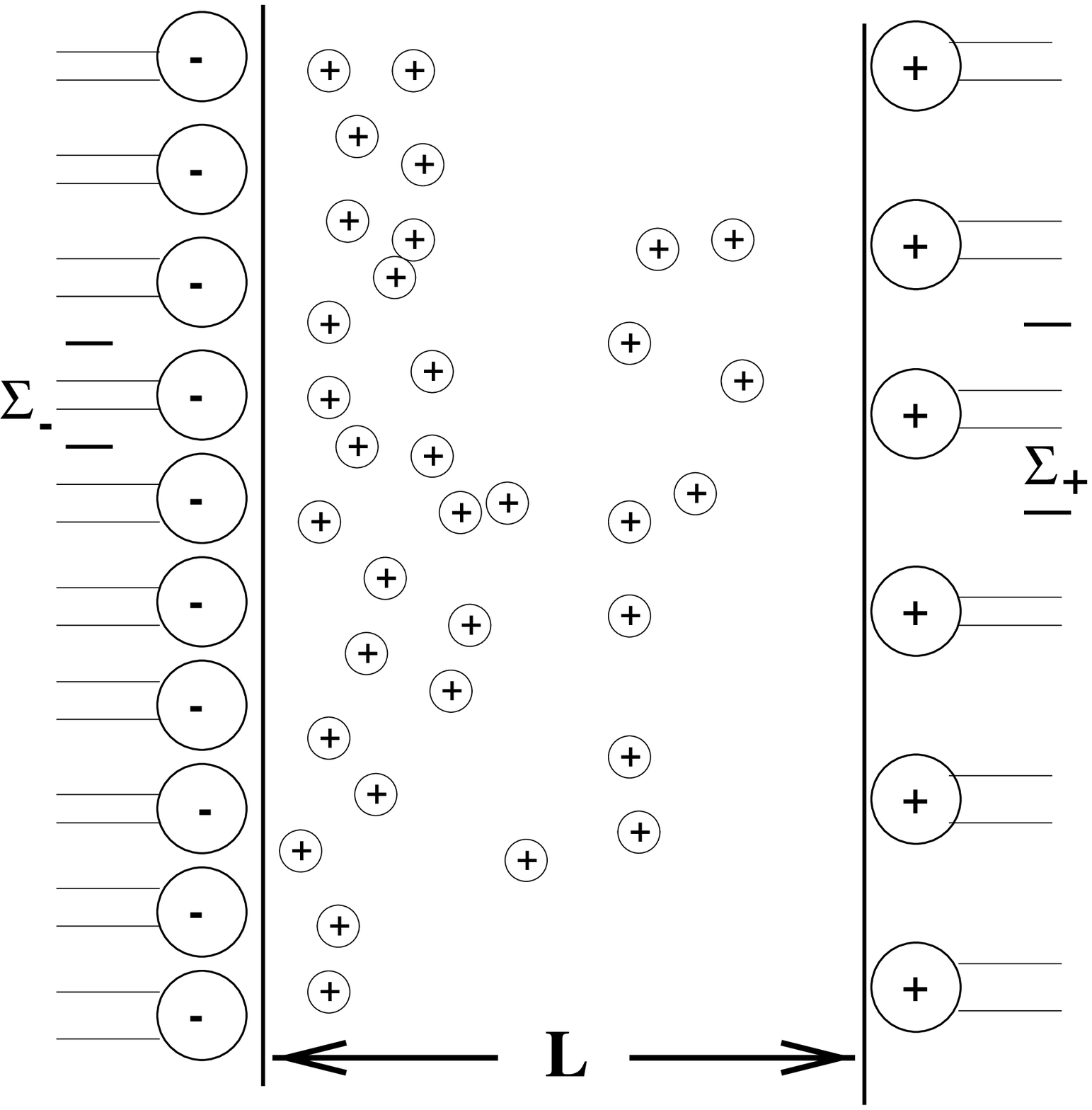}}
\vfil
\caption{Geometry of the problem: two membranes of unequal opposite
charge, separated by a distance $L$ with counterions between them.}
\label{geometry}
\end{figure}

We consider a system composed of two oppositely charged 
rigid membranes\cite{rigid} separated by a distance $L$ across an 
aqueous solution with dielectric constant $\epsilon$ 
(see Fig. \ref{geometry}).  One carries positively charged 
lipids with magnitude $q$ per head group and area per head $\Sigma_{+}.$  
The other membrane is composed of negatively charged lipids of charge 
$-q$ with $\Sigma_{-}$ area per head.  Without loss of generality, 
we assume here that $\Sigma_{-}^{-1} > \Sigma_{+}^{-1}$.  
We imagine that the two membranes, each with its own counterions which
we assume to be monovalent, initially infinitely far apart
are brought into the vicinity of each other, 
where the electrostatic interaction dominates other interactions.  
Since the counterions can gain entropy by escaping to infinity, 
we only consider the positive counterions that are required 
to neutralize the system\cite{acid}.  We also assume that there is no 
salt in the solution in to order to focus on the fundamental effects 
of the electrostatic interaction (see Conclusion for a 
discussion of the effects of added salt.)  The charged lipids in the 
membrane are modeled as a fluid with an effective surface tension $\gamma$, 
which typically has a value of $\sim 0.04\,k_{B}T/$\AA$^2$ 
for bilayers; thus in addition to the electrostatic energy, 
there is a surface free energy of the form 
$\beta F_{s}/ N = \gamma\,\Sigma$ for each membrane, where $N$ is the 
number of lipids, $\beta =1/k_B T$, $T$ is the temperature, 
and $k_B$ is the Boltzmann constant.  Note that energies are expressed
in units of $k_B T$.  We are interested in the values of 
$\Sigma^{*}_{\pm}$ which minimize the total free energy -- surface and 
electrostatic energy -- as a function of $L$ and $\gamma$.  
 
To illustrate the effect of the electrostatic interaction on the 
structure of membranes, let us consider the simplest case in which 
$\Sigma_{+}=\Sigma_{-}=\Sigma$; in this situation there are no 
counterions between the charged membranes.  The electrostatic 
contribution to the free energy can be calculated 
using Gauss' law to give\cite{free}\bb
\beta F_{el} =  N \left(\frac {2\pi l_{B} L}{\Sigma}\right),
\en
where  $l_{B}  \equiv \frac{q^{2}}{\epsilon k_{B}T} \approx 7$\AA,
is the Bjerrum length for an aqueous solution of dielectric constant 
$\epsilon = 80$ and $N$ is the total number 
of lipids in the membrane.  Minimizing $F_{s} + F_{el}$ 
with respect to $\Sigma$, we find $\Sigma^{*} = 
\left(\frac{ \pi l_{B} L} {\gamma} \right)^{1/2}.$  
Hence, the optimal area per head group depends on the square root of the
distance separating the two membranes, and for a separation of 
$ \sim 10$\AA , we find $\Sigma^{*} \sim 75$\AA$^{2}$.  
In contrast, the electrostatic free energy for an isolated charged 
membrane with its own counterions is given by\cite{pincus}\bb 
\beta F_{el} \cong  -2 N \ln{\Sigma},
\en 
to within an additive constant.  Therefore, $\Sigma^{*} 
= 2/\gamma \sim 50\,$\AA$^{2}.$  Thus, the optimal area per molecule with 
electrostatic interactions may deviate substantially from that of 
an isolated charged membrane.  Note that while the order of magnitude 
is the same in both cases, the functional dependences on the parameters 
are quite distinct.  

In order to take the counterions into account for unequally charged 
membranes, let us suppose that the negatively charged membrane 
is located at $x=0$ and the positively charged one at $x=L$.
One constraint is that of charge neutrality -- the number of 
counterions must be equal to the difference
between the number of molecules on the two membranes:\bb
\int_{0}^{L} n(x) dx =  \frac{1}{\Sigma_{-}} - \frac{1}{\Sigma_{+}}, 
\label{neutral}
\en
where $n(x)$ is the counterion density which is, in the 
mean field approximation, related to the potential 
$\phi$ by the Boltzmann factor:\bb
n(x) = n_{0}\,e^{-\beta q \phi(x)}, \label{density}
\en
where $q$ is the unit of charge and the prefactor $n_0$ is 
fixed by Eq. (\ref{neutral}).  Combining
Eq. (\ref{density}) with the Poisson equation from electrostatics  \bb 
-\nabla^{2}\phi(x) = \frac {4\pi q}{\epsilon}\,n(x), 
\en 
we arrive at the Poisson-Boltzmann (PB) equation: \bb
- \frac{d^{2}\psi}{dx^{2}} = 4\pi l_{B} n_{0} \,e^{-\psi},  \label{andy0}
\en 
where we have defined $\psi(x) \equiv \beta\,q\,\phi(x)$.

Equation (\ref{andy0}) encapsulates a mean field approach to 
the many-body problem.  
It assumes that the counterions are point-like and collectively generate
an average potential $\phi(x)$ which governs how the counterions are 
themselves distributed.  Furthermore, the PB equation, which 
neglects correlations among counterions, is valid only for sufficiently
high temperatures or low surface charge densities \cite{inter,PB}. 
Therefore, within 
the mean-field approximation, our task is to solve the PB equation
subject to the boundary conditions: \bb
\left . \frac { d\psi}{dx} \right |_{0} =  + \frac{4\pi l_{B}}{\Sigma_{-}} 
\en  
and \bb 
\left . \frac { d\psi}{dx} \right |_{L} =  + \frac{4\pi l_{B}}{\Sigma_{+}}.
\en 

In the next section, we present the 
solutions to Eq. (\ref{andy0}) and discuss the equilibrium configurations
of the counterions.  In section III we analyze the electrostatic
contribution to the free energy and pressure of the system.  In section 
IV, the phase diagram of the system is presented and followed by 
a discussion of the equilibrium value of the area per head.

\section{Nonlinear Poisson-Boltzmann Solutions}
\label{sec:solutions}

Equation (\ref{andy0}) can be solved using the ``energy'' method of
classical mechanics, where we obtain a useful constant of motion, 
defined by\bb 
E \equiv \left ( \frac{d\psi}{dx} \right )^{2}- 8 \pi l_{B} n_{0}
e^{-\psi(x)}.  \label{andy1}
\en
This constant can be physically interpreted as being proportional 
to the difference 
between the electrostatic stress and thermal pressure of the counterions
($ \sim n(x)k_{B}T$).  There are three cases to consider:
i)  $E >0$, ii) $ E = 0$, iii) $E < 0$.  

For $E>0$, we have for the normalized potential\bb 
\psi(x) =  \ln \left [ \frac {8 \pi l_{B} n_{0}}{E} \sinh^{2}
\left ( \frac{\sqrt{E}}{2}(x - x') \right ) \right ] 
\en
and the counterion distribution\bb 
n(x) = \frac{E}{8 \pi l_{B} \sinh^{2}(\frac{\sqrt{E}(x-x')}{2})}. 
\en
We choose the normalized potential to be zero at $x=0$.
This determines the 
value of $n_{0} = n(0)$. Using the boundary conditions,
one can show that the counterion densities at the surface of the membranes
are given by: \bb 
n(0) = \frac{2 \pi l_{B}}{\Sigma_{-}^{2}} - \frac{E}{8 \pi l_{B}}
\label{andy9} \en
and \bb 
n(L) = \frac{2 \pi l_{B}}{\Sigma_{+}^{2}} - \frac{E}{8 \pi l_{B}},
\label{andy10}  \en
where $E$ satisfies\bb 
E= \frac{ (4 \pi l_{B})^{2}}{\Sigma_{+}\Sigma_{-}} - \frac{4 \pi l_{B} 
(\Sigma_{+} - \Sigma{-})}{\Sigma_{+}\Sigma_{-}} 
\sqrt{E}\coth\frac{\sqrt{E}L}{2}.  \label{andy2}
\en 

For $E=0$, we have for the normalized potential\bb 
\psi(x)= \ln \left[2 \pi l_{B} n_{0} (x - x')^{2}\right]
\en
and the counterion distribution\bb
n(x) = \frac{1}{2 \pi l_{B}}(x-x')^{-2}.
\en
The boundary conditions determine the value of $L$: \bb
L = \frac {\Sigma_{+}-\Sigma_{-}}{2 \pi l_{B}}.  \label{andy3}
\en
The density at each surface can be obtained using Eqs. (\ref{andy9}) and 
(\ref{andy10}) by setting $E=0.$

For $E= - E_{-} < 0,$ we have for the normalized potential\bb 
\psi(x)= \ln \left[ \frac {8 \pi l_{B} n_{0}}{ E_{-}} \cos^{2} \left ( 
\frac{\sqrt{E_{-}}}{2}(x - x') \right ) \right ] 
\en
and counterion density\bb
n(x) = \frac{ E_{-}}{8 \pi l_{B}} \sec^{2} \frac{\sqrt{ E_{-}}(x-x')}{2}.
\label{dist}
\en
The boundary conditions give\bb
E_{-} =   \frac{4 \pi l_{B} (\Sigma_{+} - \Sigma_{-})}{\Sigma_{+}\Sigma_{-}} 
\sqrt{E_{-}}\cot\frac{\sqrt{E_{-}}L}{2} -  \frac{ (4 \pi l_{B})^{2}}
{\Sigma_{+}\Sigma_{-}}.    \label{andy4}
\en
Eqs. (\ref{andy9}) and (\ref{andy10}) are still valid 
with the replacement $E \rightarrow -E_{-}$. 

\begin{figure}
\epsfxsize = 4in
\epsfysize = 4in
\centerline{\epsfbox{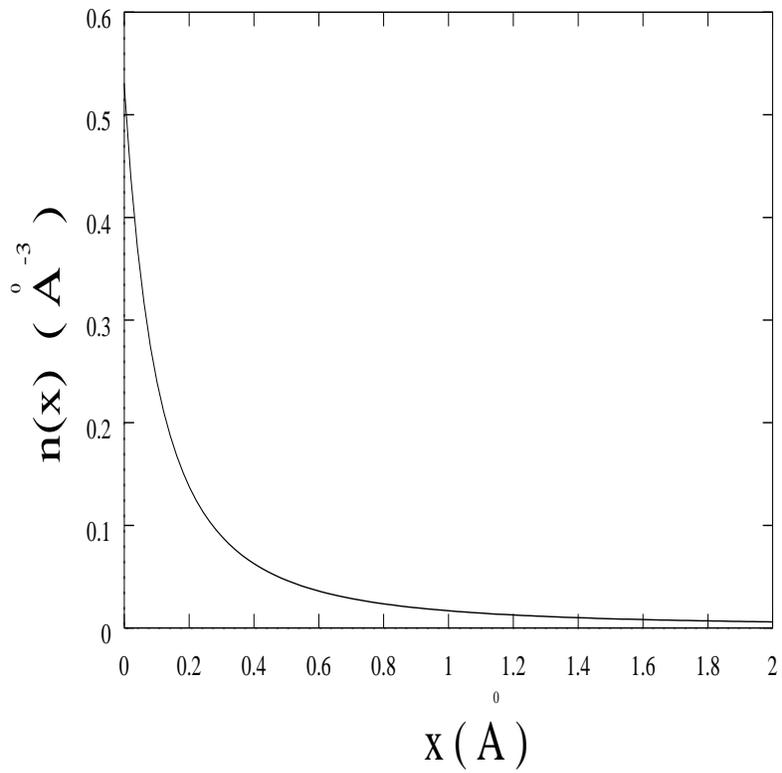}}
\vfil
\caption{The counterion distribution for the case $E < 0 $ given by
Eq. (\ref{dist}).  It is not uniform for small 
distance in contrast to the case of two
electric double layers of equal charge density of the same sign.}
\label{profile}
\end{figure}

We observe that for the dilute counterion limit, $E>0$, the equilibrium 
density distribution is essentially exponential.  However, as the
counterion density increases, the collective effect of mutual repulsion
of the counterions leads to a $\sec^2$ dependence of the counterion 
distribution as shown in Fig. \ref{profile}.  Note that
even for small separations, the counterions
density is not uniform, in contrast to the case of electric 
double layers of like charges.   

\section{The Electrostatic Free Energy and Pressure}
\label{sec:freeenergy}

\begin{figure}
\epsfxsize = 4in
\epsfysize = 4in
\centerline{\epsfbox{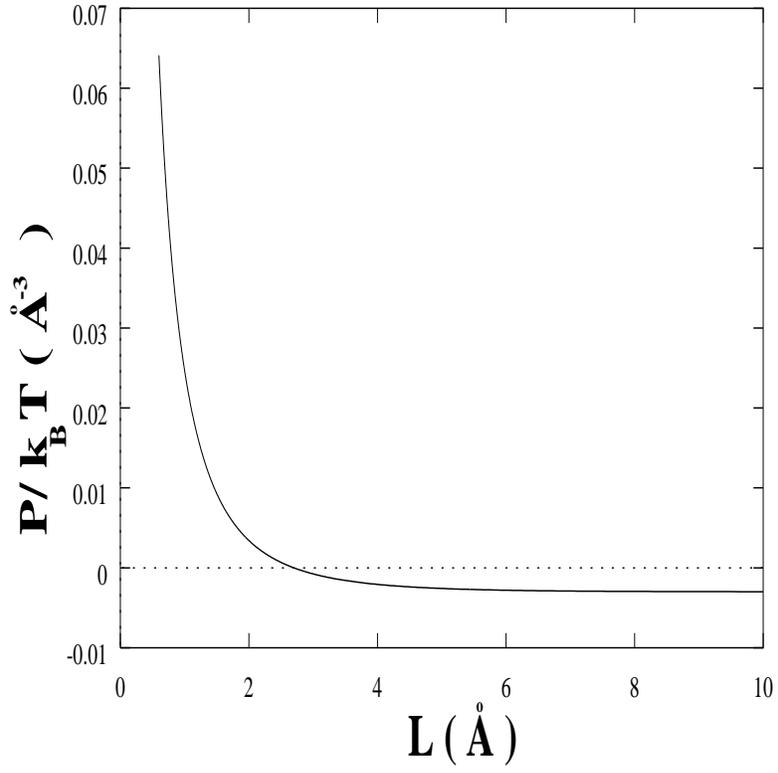}}
\vfil
\caption{A plot of pressure vs. separation distance for the 
case of $\Delta = 0.1$ and $\Sigma_{+} = 130$\AA$^{2}$ from  
Eq. (\ref{andy6}).}
\label{plot}
\end{figure}

The electrostatic free energy per unit area, which is the sum of the
electrostatic energy and the entropy of the counterions, can be written
as\bb
\beta f_{el}=  \int_{0}^{L}\,dx \left \{ \frac{E}{8 \pi l_{B}}
 -  n(x)\,\left [\,\psi(x) - \ln(n_{0}v_{0})\,\right]\, \right \},
\label{andy5}
\en
up to an additive constant, where $v_{0}$ is the volume per counterion.  
From Eq. (\ref{andy5}) or using a known expression for
the pressure \cite{PB},$$
\beta P = n(0) - ( \nabla \psi )^2 / 8 \pi l_B, 
$$
we obtain\bb
P = - \frac{k_{B}T}{8 \pi l_{B} } E.   
\label{andy6}
\en
Here we have the simple result that the pressure is proportional to $-E.$
Hence the solutions with $E>0$ and $E<0$ describe membranes that attract
and repel each other, respectively.  For the solution of $E=0$, the membranes
exert no net force on each other.  Fig. \ref{plot} shows how the pressure
varies with the distance.  Using Eq. (\ref{andy3}), we can determine 
the equilibrium distance $L^{*}$:$$
L^{*} = \frac{\Sigma_{+} - \Sigma_{-}}{2 \pi l_{B}}.
$$
Therefore, the result in Eq. (\ref{andy6}) leads to the following picture.
For large separation, $L > L^{*}$, the counterion concentration is dilute
and the electrostatic attraction dominates.  On the other hand, 
the counterions are dense when the separation is small, $L < L^{*}$.  
Hence thermal pressure dominates.  
When $L = L^{*}$ the electrostatic and thermal pressure balance,
leaving zero net pressure.  Similar results have also been obtained from 
numerical solutions to the PB equation for other systems\cite{mf}.
We note that since 
$\left .\frac{\partial^2f_{el}}{\partial L^2} \right|_{L^{*}} > 0$ as 
suggested in Fig. \ref{plot}, the system is in a stable equilibrium
at $L = L^{*}$.

Upon explicit evaluation of Eq. (\ref{andy5}) and multiplying by the area
of the membrane, the electrostatic free energy becomes:$$
\beta F_{el}/N_{+} =  - y(z)^{2} z - 2\,(\Delta^{-1} - 1 ) \ln\,z +
\Delta^{-1} \ln \left [ 1 - ( \Delta\,z\,y(z))^{2}\right]$$
\bb
- \ln \left[ 1 - ( z y(z))^{2} \right],
\label{freeen}
\en
where $y_{\pm} \equiv \frac{\sqrt{E_{\pm}}L}{2}$,
$z \equiv \frac{\Sigma_{+}}{2 \pi l_{B} L}$, and $\Delta \equiv \frac{N_{+}}
{N_{-}}$.  The function $y(z)$ is defined as\bb
y(z) = \left \{  \begin{array}{ll}
                    y_{+}  & \mbox{ if $ 0 < z < (1 - \Delta)^{-1}$} \\
                    0      & \mbox{    $z = (1 - \Delta)^{-1}$} \\
                    iy_{-} & \mbox{    $z > (1 - \Delta)^{-1}\,\,\,.$}
                 \end{array} 
        \right.
\label{pressure}
\en
Note also that $y(z) \neq 0$ satisfies\bb
zy(z) = \tanh[y(z)] + \Delta\,z\,y(z)\,\left ( 1 - z\,y(z)\,\tanh[y(z)]
\right ).    \label{andy7}
\en   

In deriving Eq. (\ref{freeen}), 
we have made an implicit assumption that the areas of the 
two membranes are the same, namely, $N_{+}\Sigma_{+} = N_{-}\Sigma_{-}$.  
The validity of this assumption is justified, since the edge effect, 
which arises when the areas of the two membranes are different, 
adds to Eq. (\ref{freeen}) a correction term of order $O(A^{-2})$.  
Therefore, in the limit of large surface areas, this contribution 
is negligible.  Note that this assumption implies that the ratio 
$\Delta \equiv \frac{\Sigma_{-}}{\Sigma_{+}}$ is fixed.  Physically, 
the asymmetric parameter $\Delta$ gives the ratio between the number
of particles in each membrane.  Equations (\ref{freeen}), 
(\ref{pressure}), and (\ref{andy7}) are the final 
results from which the equilibrium properties are derived in the next 
section.

\section{Equilibrium Properties}
\label{sec:application}

\begin{figure}
\epsfxsize = 4.0in
\epsfysize = 4.0in
\centerline{\epsfbox{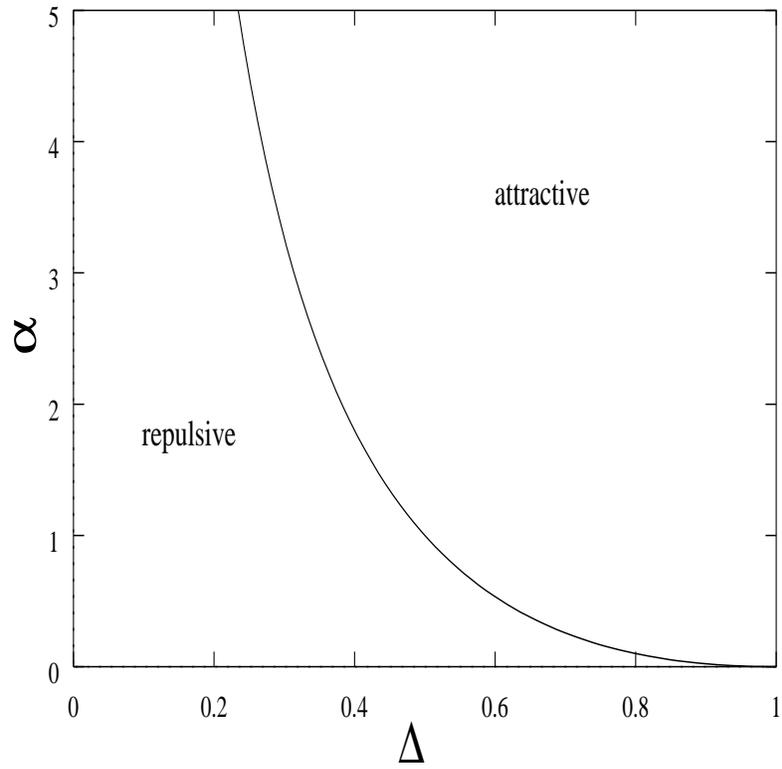}}
\vfil
\caption{The phase diagram as determined by Eq. (\ref{length}):
$\,\alpha \equiv 4 \pi l_{B} L \gamma $ vs. $\Delta$.  
The region above the line is the attractive region of 
the membranes and the region below repulsive.}
\label{phase}
\end{figure}

In this section, we determine the equilibrium value of $\Sigma_{+},$ 
using Eq. (\ref{freeen}) for the electrostatic contribution to the 
total free energy, which is given by\bb
\beta F_{tot}/N_+ = 2\gamma \Sigma_{+} + \beta F_{el}
[L, \Sigma_{+}; \Delta]/N_+,
\en
where $\gamma$ is the surface tension.  Let us first consider the case
in which the system is at equilibrium ($P = 0$).
By setting $y(z) = 0$ in Eq. (\ref{freeen}), it can be shown
that the optimal area per head group is given by\bb
\Sigma_{+}^{*} = \frac{\Delta^{-1} - 1}{\gamma},
\en
and the equilibrium distance can be calculated using Eq. (\ref{andy3}) to 
give\bb
L^{*} = \gamma^{-1}\,\,\frac{(1-\Delta)^{2}}{2 \pi l_{B}\,\Delta}.
\label{length}
\en
Note that we must restrict $\Delta$ to the range of $0 \leq \Delta < 1$,
since the case of $\Delta = 1$ corresponds to zero counterion density
where the membranes attract for all separations.  For the
case of $\Delta = 0.1$, we find $\Sigma_{+}^{*} \sim 200$\AA$^{2}$ and 
$L^{*} \sim 5$\AA.  Therefore, the counterions have yet a stronger effect 
on the stretching of the membrane compared to the single charged membrane
case where we have shown $\Sigma \sim 50$\AA$^{2}$.  
According to the discussion of Section \ref{sec:freeenergy}, 
Equation (\ref{length}) 
represents a line which separates the repulsive region and the 
attractive region in the phase diagram, which is shown 
in Fig. \ref{phase}.

\begin{figure}
\epsfxsize = 4.0in
\epsfysize = 4.0in
\centerline{\epsfbox{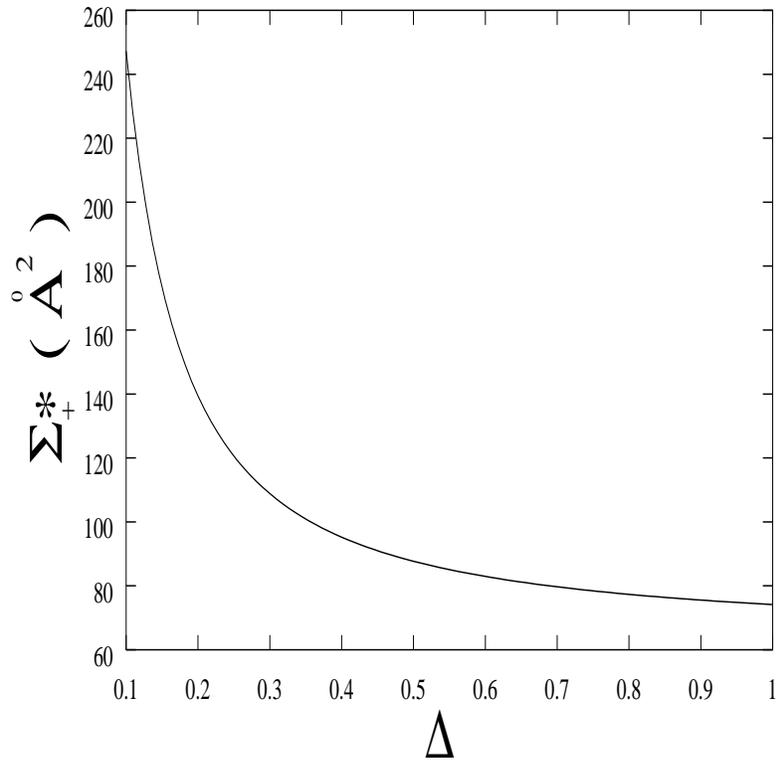}}
\vfil
\caption{The optimal value of the area per headgroup $\Sigma_{+}^{*}$
as a function of the charge asymmetric parameter $\Delta$ for $L = 10$ \AA.}
\label{sigma}
\end{figure}

Next, we consider two membranes separated by a fixed distance.
Equation (\ref{freeen}) can be regarded as a function of only $z$ with 
the help of Eq. (\ref{pressure}).  Thus, we can imagine solving $y(z)$ 
in terms of $z$, and by substituting the 
result into Eq. (\ref{freeen}), minimization of the free energy can be carried 
out explicitly.  Unfortunately, this cannot be done analytically 
in general. However, our estimate of $L^{*}$ above
indicates that for $L < L^{*}$, which is of atomic size, 
other effects such as thermal fluctuations and van der Waals attraction 
that have not been taking into account, may become significant.  
Therefore, we focus on the large separation limit, 
$L >> L^{*}$, where $ y(z) >> 1.$  With this approximation,  
we obtain\bb
\Sigma_{+}^{*} \simeq \frac{1}{2 \Delta \gamma}\left 
\{ 1 - \Delta + \left( ( 1 - \Delta)^{2}  + 4 \pi l_{B} 
\Delta^{2}\gamma L \right)^{1/2} \right \}.
\label{asig}
\en
Here, we observe that $\Sigma_{+}^{*}$ decreases as $\Delta \rightarrow 1$
monotonically as shown in Fig. \ref{sigma}.
This is a reflection of the fact that the counterions 
have a strong effect on the stretching of the membrane.  The presence of
the counterions enhances the electrostatic energy; therefore,
lowering of the charge density is energetically favorable. Hence 
the stretching of the membrane.  For $\Delta = 0.1$ and $L = 10$\AA, we
find $\Sigma_{+}^{*} \sim 250 $\AA$^{2}$ which is five times the value 
of an isolated charged membrane!

\begin{figure}
\epsfxsize = 4.0in
\epsfysize = 4.0in
\centerline{\epsfbox{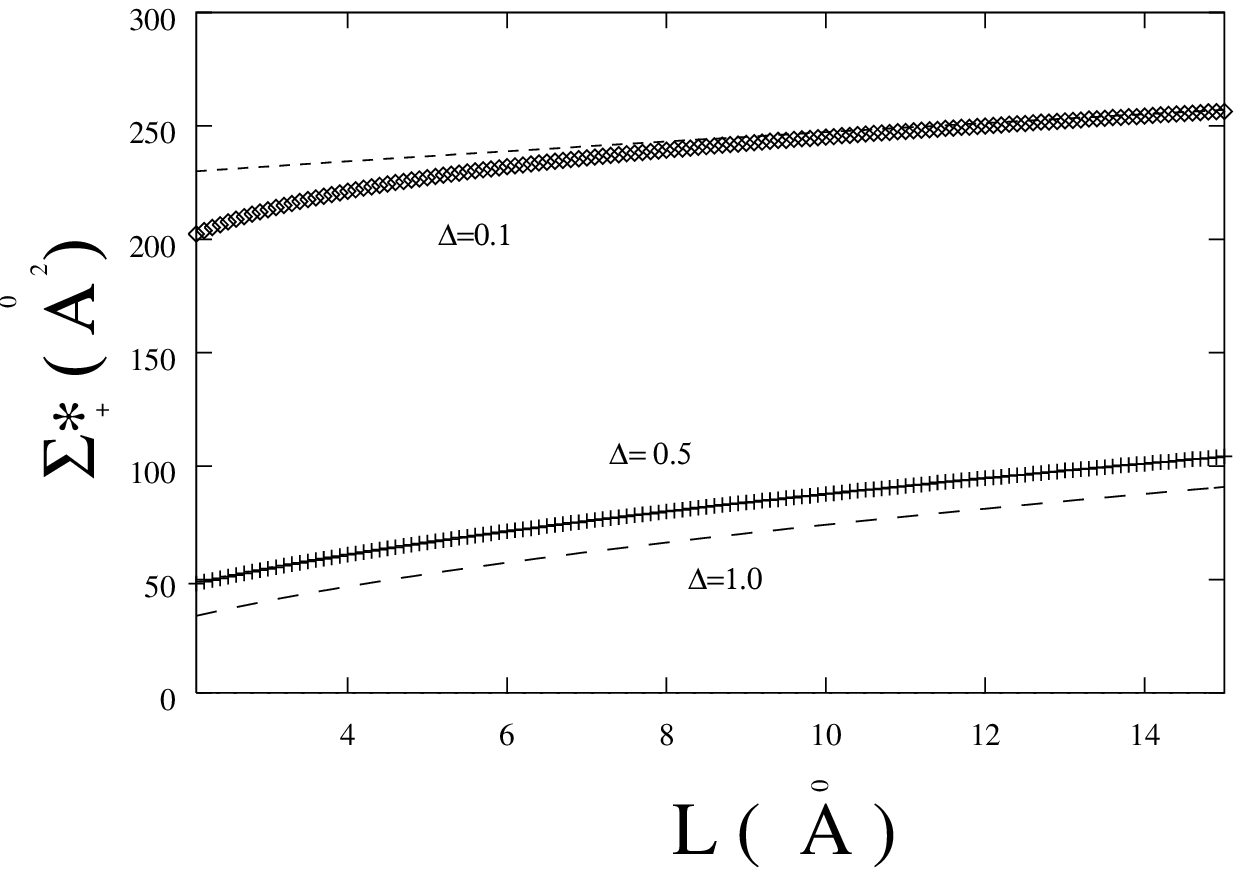}}
\vfil 
\caption{The optimal value of the area per headgroup $\Sigma_{+}^{*}$
as a function of distance, obtained numerically for $\Delta$ = 0.1, 0.5.
The curves are plots of  Eq. (\ref{asig}) for their 
respective values of $\Delta$.}
\label{sigma2}
\end{figure}

For completeness, we have performed the minimization of the free
energy numerically for physically relevant separations ($L > 2 $\AA).
The results for the cases $ \Delta = 0.1, 0.5$ are shown in 
Fig. \ref{sigma2}.  We see that Eq. (\ref{asig}) is indeed not
a bad approximation for large distances.  Furthermore, we observe that 
$\Sigma_{+}^{*}$ increases monotonically with distance for all 
values of $\Delta$, as also suggested by Eq. (\ref{asig}).   Note that
even for $L < L^{*} $, $\Sigma_{+}^{*}$ bears the same qualitative 
dependence on $\Delta$.

\section{Conclusion}
\label{sec:conclusion}
In this paper, we have considered two asymmetrically charged membranes 
of opposite sign.  The counterion distribution is obtained by solving
the PB equation.  The membranes can either attract, repel, or exert no
force on each other, depending on the difference in the area per head 
group of the membranes.   We have also made an attempt to understand
the collective effects of the counterions on the structure of 
the membranes.  Due to the electrostatic interaction of the counterions
and the membranes, the stretching of the membranes becomes energetically
favorable.  In the high counterion concentration limit ($\Delta << 1$)
the area per head can reach a value few times as large as an isolated
charged membrane.  We note here that since the volume of a 
lipid bilayer is fixed, due to the incompressibility of lipid molecules, 
the stretching of the membrane implies a decrease in its thickness.
This effect may be observed in real experimental settings 
by {\em e.g.} x-ray scattering.

Finally, we wish to comment on the relevance of the model 
studied in this paper to other systems.  First, for the
case of permeable membranes, where counterions may freely 
permeate behind the charged membranes, the electrostatic 
interaction can be shown to be attractive for all separations, 
since the thermal pressure exerted on the membrane 
by the counterions is much weaker.
In this case, we have verified that this problem is almost 
identical to our model for the solution $E>0$.  Therefore,
the renormalization of area per head group as given by Eq. (\ref{asig})
should hold at least qualitatively.  Secondly, we also consider 
the presence of a finite amount of salt in our system.  
This problem may be described at the PB level by 
an equation similar to Eq. (\ref{andy0}), $$
- \frac{d^{2}\psi}{dx^{2}} = \kappa^2 \sinh \psi + 
4\pi l_{B} n_{0} \,e^{-\psi},
$$
where $\kappa$ is the inverse of the screening length.  The above
equation can still be solved in principle but involves rather complicated 
mathematical functions.  This is a subject for further study.  
However, it can be inferred that our model
corresponds to the case where the separation between 
the membranes is small compared to the screening length, 
{\em i.e.} $L < \kappa^{-1}$.  Typically, for 
$\kappa^{-1} \sim 20 $\AA$\,$ for 0.05 M of salt, our model should 
provide a good picture for the binding of oppositely charged membranes 
for a distance of the order of few \AA.  For $\kappa^{-1} \geq L$, 
the electrostatic interaction is exponentially screened.  
Nevertheless, according to Ref. \cite{unequal} 
even with a finite amount of salt, the {\em linearized} PB theory 
also predicts regions of repulsion and attraction between membranes,
similar to what we have found here.  Therefore, we believe our model 
captures the essential physics associated with the 
electrostatic interaction of oppositely charged membranes.    

\section{Acknowledgments}
\label{acknowledgments}

We would like to thank R. Bruinsma, V. A. Parsegian, W. Gelbart,
and C. Safinya for stimulating and helpful discussions.
AL and PP acknowledge support from NSF grants MRL-DMR-9632716, 
DMR-9624091, DMR-9708646, UC-Biotechnology Research and Education 
Program (97-02), and US-Israel Binational Science Foundation.

\pagebreak

\newpage

\newpage

\newpage

\newpage

\newpage


\begin{thebibliography}{99}

\bibitem{inter} 
J. N. Israelachvili, {\em Intermolecular and Surface Forces.}
(Academic Press Inc., San Diego, 1992); B. Honig and A. Nicholls,
Science {\bf 268}, 1144 (1995).

\bibitem{unequal}
V. A. Parsegian, Gingell D., Biophys. J. {\bf  12}, 1192 (1972).

\bibitem{biology}
B. Alberts, D. Bray, J. Lewis, {\em Molecular Biology of the Cell.} 
(Garland Publishing, New York, 1994).

\bibitem{gene}
R.G. Crystal, Science {\bf 270}, 404 (1995).


\bibitem{adhesion}
J. Nardi, T. Feder, R. Bruinsma, and E. Sackmann, Europhys. Lett. 
{\bf 37}, 371 (1997).

\bibitem{DNA}
J. O. Radler, l. Koltover, T. Salditt, C. R. Safinya, Science {\bf 275},
810 (1997).

\bibitem{con}
J. H. Schulman and J. B. Montagne, Ann. N. Y. Acad. Sci. {\bf 92}, 366 (1961);
P. G. de Gennes and C. Taupin, J. Phys. Chem. {\bf 86}, 2294 (1982).

\bibitem{rigid}
For a discussion on the stiffening of charged membranes, see 
D. Andelman, in {\em Handbook of Biological physics}, 
edited by R. Libowsky and E. Sackman 
(Elsevier Science Publishers, Amsterdam, 1995) and references therein. 

\bibitem{acid}
It should be pointed out that the justification to this statement 
could be quite subtle and is an area for further study.  However,
consider an experimentally relevant scenario where the positively 
charged membrane is composed of acidic lipids and negatively charged 
membrane basic.  The counterions released by the membranes are 
H$^{+}$ and OH$^{-}$.  Therefore, assuming the system has had time
to equilibrate, the counterions eventually form H$_{2}$O 
upon recombination and can be considered as escaping to infinity.  

\bibitem{free}
Note that this expression is infinite for infinite separation.  
Thus, we have to add to this expression a distance dependent term,
in association with the entropy gained by the counterions.  However,
this constant does not play any role in what follows.   

\bibitem{pincus}
P. Pincus, in {\em Phase Transitions in Soft Condensed Matter}, edited
by E. Riste and D. Sherrington (Plenum Publishing Corporation, 1989).

\bibitem{PB}
S. A. Safran, {\em Statistical Thermodynamics of Surfaces, Interfaces,
and Membranes} (Addison-Wesley Publishing Com., Reading, 1994).

\bibitem{mf}
N. Ben-Tal, B. Honig, R. M. Peitzsch, G. Denisov, and S. McLaughlin, 
Biophys. J. {\bf  71}, 561 (1996); N. Ben-Tal, B. Honig, C. Miller,
and S. McLaughlin, Biophys. J. {\bf  73}, 1717 (1997).

\end{thebibliography}
\end{document}